# Simplified Design of Optical Elements for Filled-Aperture Coherent Beam Combination

CHRISTOPHER ALESHIRE[1,*], ALBRECHT STEINKOPFF[1], CESAR JAUREGUI[1], ARNO KLENKE[1,2], ANDREAS TÜNNERMANN[1,2,3] AND JENS LIMPERT[1,2,3]

[1]*Institute of Applied Physics, Friedrich-Schiller University Jena, Albert-Einstein-Straße 15, 07745 Jena, Germany*
[2]*Helmholtz-Institute Jena, Fröbelstieg 3, 07743 Jena, Germany*
[3]*Fraunhofer Institute for Applied Optics and Precision Engineering, Albert-Einstein-Straße 7, 07745 Jena, Germany*
*\*christopher.aleshire@uni-jena.de*

**Abstract:** A simplification strategy for Segmented Mirror Splitters (SMS) used as beam combiners is presented. These devices are useful for compact beam division and combination of linear and 2-D arrays. However, the standard design requires unique thin-film coating sections for each input beam and thus potential for scaling to high beam-counts is limited due to manufacturing complexity. Taking advantage of the relative insensitivity of the beam combination process to amplitude variations, numerical techniques are used to optimize highly-simplified designs with only one, two or three unique coatings. It is demonstrated that with correctly chosen coating reflectivities, the simplified optics are capable of high combination efficiency for several tens of beams. The performance of these optics as beamsplitters in multicore fiber amplifier systems is analyzed, and inhomogeneous power distribution of the simplified designs is noted as a potential source of combining loss in such systems. These simplified designs may facilitate further scaling of filled-aperture coherently combined systems.

## 1. Introduction

By coherently combining the outputs of multiple laser sources, the fundamental and practical limitations of single-emitter laser sources can be overcome [1]. Experiments with Coherent Beam Combination (CBC) have at present demonstrated the scaling of average powers by about one order of magnitude in the highest power experiments [2]. These demonstrations and the further scaling potential of CBC have made it a promising technology for the development of next-generation, ultra-high-power laser sources in major research facilities [3]. While these proposed systems are promising for the achievement of new scientific objectives, their practical realizations will be truly one-of-a-kind systems requiring significant investment. Even for contemporary high-power CBC systems, scaling of performance by an order of magnitude requires a substantial financial investment and a significant allotment of laboratory space. In standard CBC systems the practical requirements scale more or less linearly with the increase in power: a system with *N*-times as much power will require *N*-times as much space in a laboratory and *N*-times as much financial investment. This scaling law makes increasing performance by multiple orders of magnitude largely impractical and presents a major hurdle for the maturation of CBC into a mainstream laser technology.

In the case of coherently combined fiber amplifier systems, the above-mentioned scaling law can be circumvented by the use of multicore fibers (MCFs) [4]. In such fibers, multiple doped cores are arranged in a single shared pump cladding, allowing a potential *N*-fold power scaling from a single fiber. Recent experiments have demonstrated kilowatt power levels and the combination of femtosecond pulses in relatively compact systems [5]. The space efficiency of these systems is aided by the use of multi-channel devices including piezo- and photodiode arrays for phase detection and actuation, and Segmented Mirror Splitters (SMS) for beam division and recombination. In particular, SMS elements significantly reduce optical path lengths and the number of optical components required for filled-aperture beam combining. These elements replace a cascaded series of beamsplitters for splitting a single beam into a linear array. As beam combining and beamsplitting are spatially and temporally symmetric processes, the same optics can be used in an inverse configuration with phase-matching to recombine a linear array into a single beam (Fig. 1a). Similar devices have been proposed for pulse stacking applications [6] and have been demonstrated for beam division and recombination with intracavity passive CBC [7]. With two SMS optics these functions can be extended to a 2-D array, for example to distribute a seed laser to the cores of an MCF amplifier with a rectangular core arrangement or to recombine its amplified output beams. An SMS can be manufactured with only one or two optical elements, as shown schematically in Fig. 1. For combination of ultrafast lasers, a two-element air-spaced design is preferable to minimize nonlinear and dispersive effects (Fig. 1b).

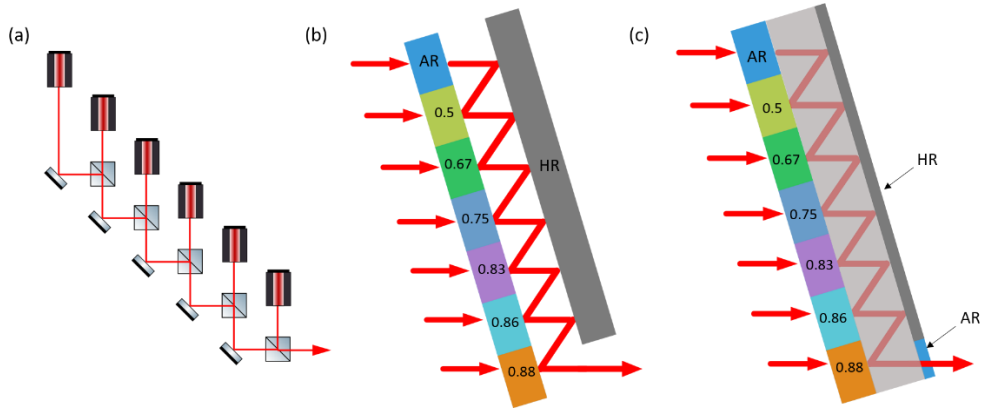

Fig. 1. (a) Schematic representation of filled aperture CBC with cascaded beamsplitters. (b) Equivalent two-element SMS. Numbers indicate the reflectivities of the beam-splitting surfaces which give perfect amplitude matching, assuming uniform powers in the input beam array. HR and AR indicate highly-reflective and anti-reflective surfaces, respectively. (c) Equivalent monolithic SMS, where all coatings are deposited onto a single optic.

While SMS optics provide a compact means for division and combination of beams, in their standard design these devices are not easily scalable to high beam counts. For perfect amplitude matching, a unique coating section must be used at each splitting or combining junction. For example, a ten-beam SMS will require ten adjacent coatings with unique, tightly-toleranced reflectivities deposited onto a single optical surface, as shown in Fig. 1b. This design requirement makes the design and manufacture of high-beam count SMS optics prohibitively complex.

However, it can be shown that the effect of amplitude error in coherent combination is relatively small in comparison to that of phase error [8]. Thus, the resilience of the coherent process to amplitude errors can be exploited in the design of simplified high-beam count SMS optics based on a reduced number of coating sections. In this article, a general SMS design principle and three numerically simulated versions thereof are described. The potential performance of the simplified designs as both beam combiners and beam splitters is discussed, and a specific analysis with respect to coherently combined MCF amplifiers is given.

## 2. SMS Simplification for Coherent Beam Combination

The SMS simplification concept relies on using a reduced number of beam-splitting coatings, the reflectivities of which are determined by numerical optimization of the coherent combination process. Three example designs are given in Fig. 2. In every design variant, the first input beam is transmitted through an anti-reflective (AR) coating, while the remaining $N$-1 beams are incident on and divided among one, two, or three coating sections with reflectivities determined by the optimization algorithm. These design variants are here referred to as a "1-Coating," "2-Coating," and "3-Coating" SMS, respectively. The combining efficiency, $\eta$, achieved at each combining step in a simulated SMS can be calculated as in the case of two-beam interference of two plane-waves incident on orthogonal ports of a beamsplitter with reflectivity $R$:

$$\eta = \frac{RI_1 + (1-R)I_2 + 2\sqrt{R(1-R)I_1 I_2}\cos(\Delta\phi)}{I_1 + I_2} \quad (1)$$

When applying this equation to an SMS, $I_1$ is the power of the field travelling within the SMS, $I_2$ is the power of the next beam in the linear array to be combined and $\Delta\varphi$ denotes the relative phase difference between the beams. Throughout the simulations described here, perfect phase-matching is assumed, reducing the cosine term to unity in all cases. Identical input beam powers are also assumed, as this would generally be the case in a high-power system where parallel emitters are operated near their practical or physical limitations. By iteratively applying Eq. 1, the theoretical combining efficiency of a given SMS design can be calculated. With the overall combining efficiency as a merit function, numerical optimization yields the ideal simplified SMS design parameters. For a 1-Coating SMS, only the reflectivity of the single shared coating section is optimized. In the 2-Coating and 3-Coating designs, the reflectivities of the unique coating sections are optimized in addition to the number of beams incident on each coating section, $N_1$, $N_2$, and $N_3$, referred to here as the "coating populations."

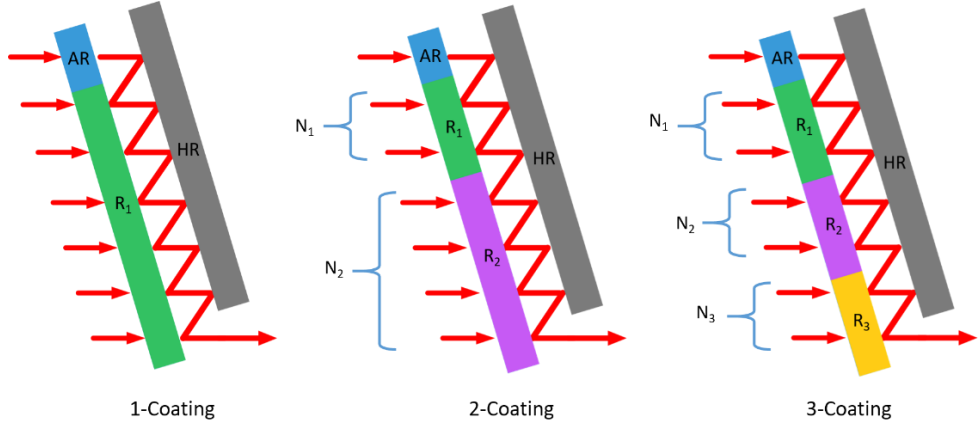

Fig. 2: Three simplified SMS design variants. $R_1$, $R_2$, $R_3$ indicate reflectivities of coating sections shared between multiple input beams. $N_1$, $N_2$, $N_3$ indicate the number of beams shared in each coating section.

The optimized theoretical combining efficiencies of the three design variants for combination of up to forty beams in a single SMS are shown in the upper graph of Fig. 3. In these simulations the HR and AR surfaces are assumed to be ideal in order to show only the residual effect of amplitude mismatch in the optimized designs. The solid black line indicates the ideal combining efficiency of a standard SMS design where a unique coating is used for every input beam to achieve perfect amplitude matching, which is always 100% with ideal AR and HR surfaces. In practice, however, these surfaces will not function ideally, and this can be easily integrated into the numerical optimization process. The greatest effect will arise from a non-ideal HR surface, as $N$-1 reflections will occur on this surface, causing an unavoidable loss which is also present in the standard SMS design. The lower graph in Fig. 3 shows the optimized combining efficiencies of the three simplified SMS variants with a 99.8% reflective HR surface. For thin-film coatings in the visible or near infrared, this reflectivity is quite low, but gives a pessimistic baseline for the potential losses. The solid black line indicates the combining efficiencies of the standard SMS design with the same HR reflectivity. Unsurprisingly, as more coatings are incorporated into a simplified design, a higher combination efficiency can be achieved as the optimized coating sections can more closely approximate the ideal amplitude-matching scenario. The simplified design concept is, of course, not limited to the three simplest variations shown in Fig. 2, as additional coatings can be added to balance manufacturing complexity with achievable combining efficiency. However, the marginal benefit of adding additional coating sections diminishes quickly. For example, using only three unique coatings for the combination of forty beams in a single SMS results in only 2% additional combining loss compared to the standard SMS design, which would require thirty-nine unique beamsplitter coatings.

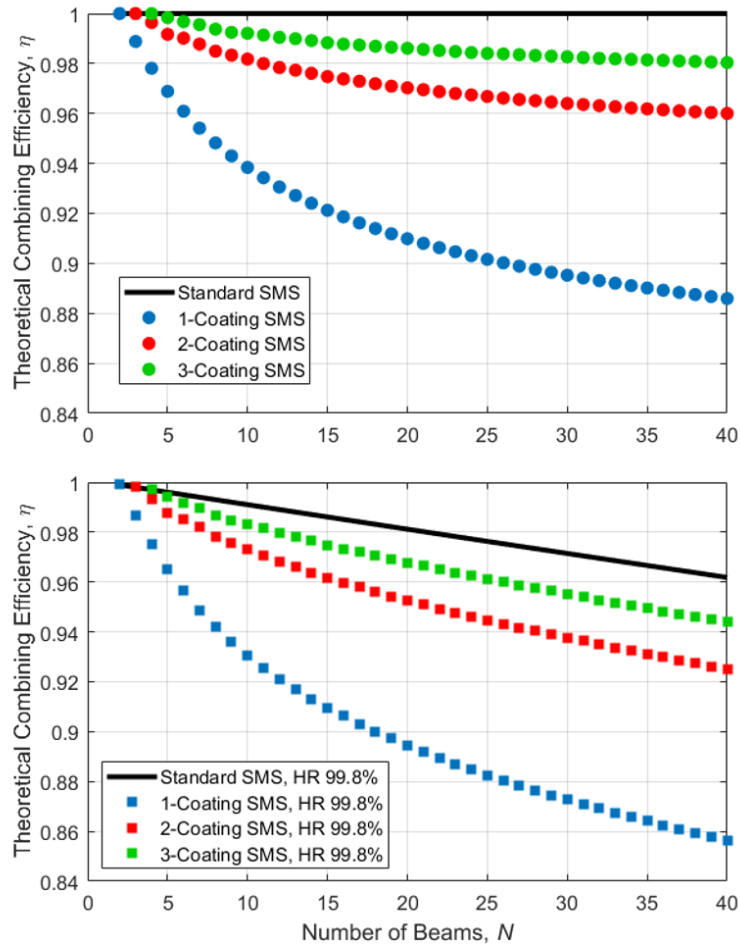

Fig. 3: Optimized combining efficiencies of the three simulated SMS designs. (Top) Theoretical combining efficiencies assuming ideal AR and HR surfaces. (Bottom) Theoretical combining efficiencies with 99.8% reflectivity HR surface. In each plot, the solid black line indicates the efficiency achievable with the standard SMS design as shown in Fig. 1

As mentioned previously, in the optimized 2-Coating and 3-Coating SMS designs the coating populations are also optimized in addition to the coatings themselves. This is not a strict requirement, and the coating populations may be held to some fixed number or limited span in order to restrict the real dimensions of the coating sections as required for manufacturing. This will only slightly reduce the expected combining efficiency. Nevertheless, unrestricted optimization of the coating populations yields the highest possible efficiency. The optimized coating reflectivities and the corresponding optimized coating populations are shown in Fig. 4 for the 2-Coating and 3-Coating designs. These optimized parameters and all further analysis are for the case of an ideal HR surface. In the case of a non-ideal HR surface, the optimized coating populations and ideal reflectivities will differ slightly, particularly for higher channel counts. Again, non-ideal parameters can be easily integrated into the optimization algorithm to account for real-world parameters. In both designs shown in Fig. 4, the later coating populations grow at a faster rate, as the relative difference in amplitude mismatch between adjacent beam combining steps is smaller in the later combining stages and, thus, can be more easily compensated for with multiple beams in a single coating than in the earlier combining steps. Note that the optimized coatings of the 1-Coating design have been omitted as the efficiencies at higher channel counts are relatively low (see Fig. 3).

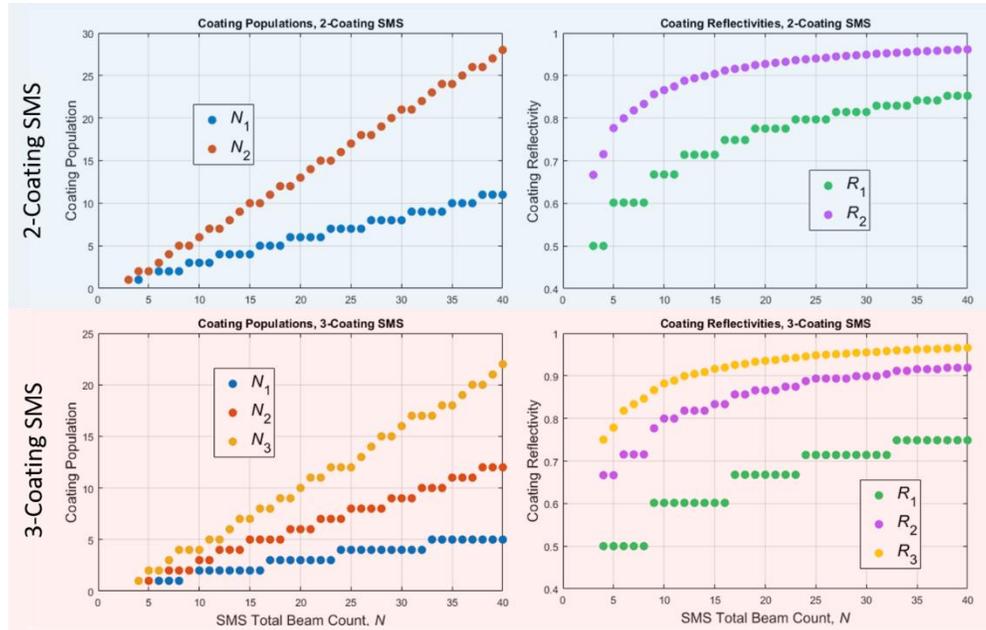

Fig. 4: Optimized parameters for 2-Coating and 3-Coating SMS designs. The blue and red sections contain plots of the optimum parameters for 2-Coating and 3-Coating SMS designs, respectively. Within each colored section the right plot gives optimized coating populations and the left plot contains optimized reflectivities of the coating sections.

As the combining process is relatively insensitive to amplitude errors, the beam combining efficiency of the simplified designs is also tolerant to small deviations from the optimum coating reflectivities. To demonstrate this, the combining efficiency of the simplified designs can be tested with an error in coating reflectivities. A full analysis of all coating error scenarios is beyond the scope of this article. As an example, the effect of coating error for the 1-Coating SMS design is shown in Fig. 5. This is the simplest error case as there is only a single optimized coating and, thus, the resistance to error can be easily illustrated. A positive error in reflectivity ranging from one to four percent is applied to the optimized coatings for up to twenty-five beams, and combination efficiency is tested in each case. In this scenario a positive error always produces a greater additional loss than a negative error of the same magnitude, as a positive error will more dramatically impact amplitude matching in the case of greater power in the beam which is reflected in combination ($I_1$ in Eq. 1). Therefore, this plot represents the worst-case loss scenario. This plot is restricted to twenty-five beams to show the smaller error cases in more detail, as the four percent error case diverges quickly for higher channel counts.

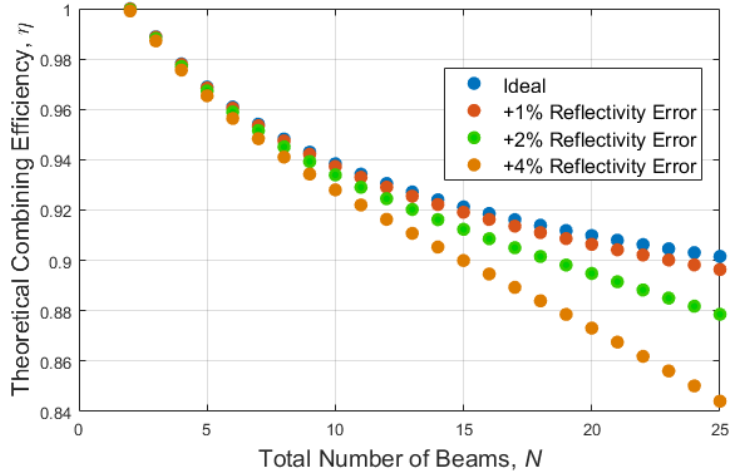

Fig. 5: Effect of coating error in 1-Coating SMS design. A positive reflectivity error of up to 4% is applied to the ideal optimized coatings. A positive error was found to have the greatest detrimental effect in this scenario, thus representing the worst-case scenario.

The simulations and analysis presented here were conducted in order to examine the effect of amplitude mismatch in SMS beam combiners and to understand how this effect can enable simplified designs based on a compromise between device complexity and the amplitude matching requirement. The results indicate that a high combination efficiency can be achieved for several tens of beams with highly-simplified SMS designs. These simulations do not consider a number of other effects which may be of interest to those wishing to use SMS devices as beam combiners. For example, combining efficiencies have been determined from a plane wave analysis, and losses due to spatial mismatch between beams may be induced by e.g. poor surface quality or coating-induced wavefront errors. The loss caused by such mechanisms has been previously studied elsewhere [9]. As these loss mechanisms are strongly dependent on factors other than the SMS design, such as manufacturing capabilities and laser system design, they have been omitted from the analysis of this article.

## 3. Beam Division with Simplified SMS Designs

The simulations presented above indicate that a high combining efficiency can be achieved with simplified SMS designs used as beam combiners. Additionally, as SMS optics are also used as beamsplitters for seeding the array of cores in MCF amplifiers (see [4]), it is also important to consider the potential performance of these simplified designs as beamsplitters. Just as in combining, perfect uniformity of beams emerging from an SMS used as a beamsplitter can only be achieved with the standard SMS design utilizing a unique coating for each beam division step. However, while the coherent process of beam combining is resistant to amplitude errors, the incoherent process of beam-splitting does not benefit from this effect. It is important to note that numerical optimization of a reduced number of coatings for combining efficiency as in Section 2 or for uniformity of beam-splitting will yield the same solutions. Optimization for combining efficiency using Eq. 1 minimizes the cumulative effect of amplitude error in the coherent summation of the input fields, and, therefore, the optimization of the power (amplitude) splitting uniformity will return an identical result for coating reflectivities and coating populations.

It is straightforward to computationally test the splitting uniformity of any of the optimized designs described in Section 2. A useful metric to consider is the ratio of the minimum and maximum output beam powers from an SMS used as a beamsplitter. Fig. 6a shows this ratio for optimized 3-Coating SMS devices for five to twenty-five beams. For each channel count, a 3-Coating SMS design provides the best amplitude homogenization in splitting, just as the 3-Coating SMS used as a combiner can best approximate the ideal amplitude-matching scenario and achieve the highest combining efficiency. As an example, Fig. 6b shows the normalized splitting of a ten beam 3-Coating SMS. The uniformity of beam-splitting with SMS optics is particularly important in the case of MCF amplifier systems, where a number of optical processes will be influenced by the absolute seed power in each core and by the relative seed powers which will in turn affect the uniformity and combinability of the amplified beams. Using the ten beam SMS in Fig. 6b as an example, in a 10x10 MCF amplifier seeded by an array formed by two of these optics, the ratio between minimum and maximum seed powers will be 0.28. In the case of cw MCF amplifier systems, such non-uniform seed light distribution may be tolerable in general, provided that the overall seed power is adjusted so that all cores are seeded in the "deeply-saturated" regime. In this case, and with a high overall gain coefficient, the amplified beams exiting the MCF cores will have only small variations in output power and in optical signal-to-noise ratio, yielding little loss in the subsequent combining process. In the case of ultrafast MCF amplifiers, however, a non-uniform distribution of seed light will lead to core-to-core variation in accumulated nonlinear phase (B-integral) through the amplification process. In fibers this variation is exacerbated by high peak power in fiber cores and long path lengths. Thus, an analysis of the effect of non-uniform beam splitting with simplified SMS optics in MCF amplifiers is highly dependent on the parameters of the fiber system.

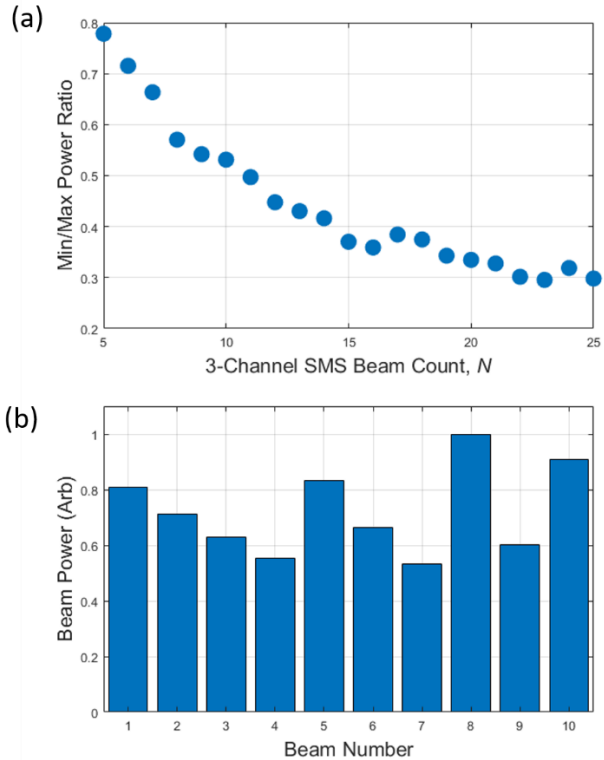

Fig. 6: Beam-splitting with optimized 3-Coating SMS designs. (a) Ratios of minimum to maximum output powers of 3-Coating SMS designs from 5 to 25 beams. (b) Histogram of output beam powers from a 10-beam 3-Coating SMS used as a beamsplitter.

An example of an analysis to determine suitability of a simplified SMS as a beamsplitter is given in Fig. 7. A fiber amplifier simulation tool was used to model amplification in a one-meter long single-core double-clad Yb-doped fiber with a core diameter of 30 microns. The modelled amplifier is counter-pumped with 100 W at 976 nm and seeded with a range of average powers from 2 W to 30 W at wavelength 1030 nm. The power evolution along the fiber is calculated for each seed power, leading to the distributions plotted in the upper graph of Fig. 7. Note that the power evolution along the fiber is nearly identical for each seed power, indicating that all seed levels are saturating the amplifier gain. For every tested seed power, a B-integral can be calculated by integrating over this plot with the standard formulation [10]:

$$B = \frac{2\pi}{\lambda} \int n_2 I(z) dz \qquad (2)$$

Where $z$ is the distance along the fiber, $\lambda$ is the signal wavelength, $n_2$ is the nonlinear refractive index (here taken to be $n_2 = 3 \cdot 10^{-20}$ W/m² for fused silica), and $I$ is the average intensity in the fiber core at each point along the fiber (calculated as the average power divided by the effective mode area.) With the calculated B-integrals, a relationship between average seed intensity and B-integral can be plotted, as shown in the lower graph of Fig. 7. The linearity of this data again signifies that the tested seed powers are in a well-saturated regime of the amplifier operation. A fitted linear slope can be made to determine the variation of the B-integral with respect to a variation in seed intensity, in this case yielding $\frac{dB}{dI} = 1.6 \cdot 10^{-13} \frac{rad \cdot m^2}{W}$. This formula can be extended to incorporate the peak power in the case of a pulsed system by integration of the pulse duration and repetition rate. Then the difference in B-integral in this amplifier for different power levels of a pulsed seed source is calculated as

$$\Delta B = 1.6 \cdot 10^{-13} \cdot (0.94) \cdot \left(f_{rep} \cdot \tau_{pulse}\right)^{-1} \cdot \Delta I_{avg} \qquad (3)$$

Where $f_{rep}$ is the seed laser repetition rate, $\tau_{pulse}$ is the pulse duration, and $\Delta I_{avg}$ is the difference in intensity calculated with average power. A factor of 0.94 is included for the calculation of peak power with Gaussian pulse shape.

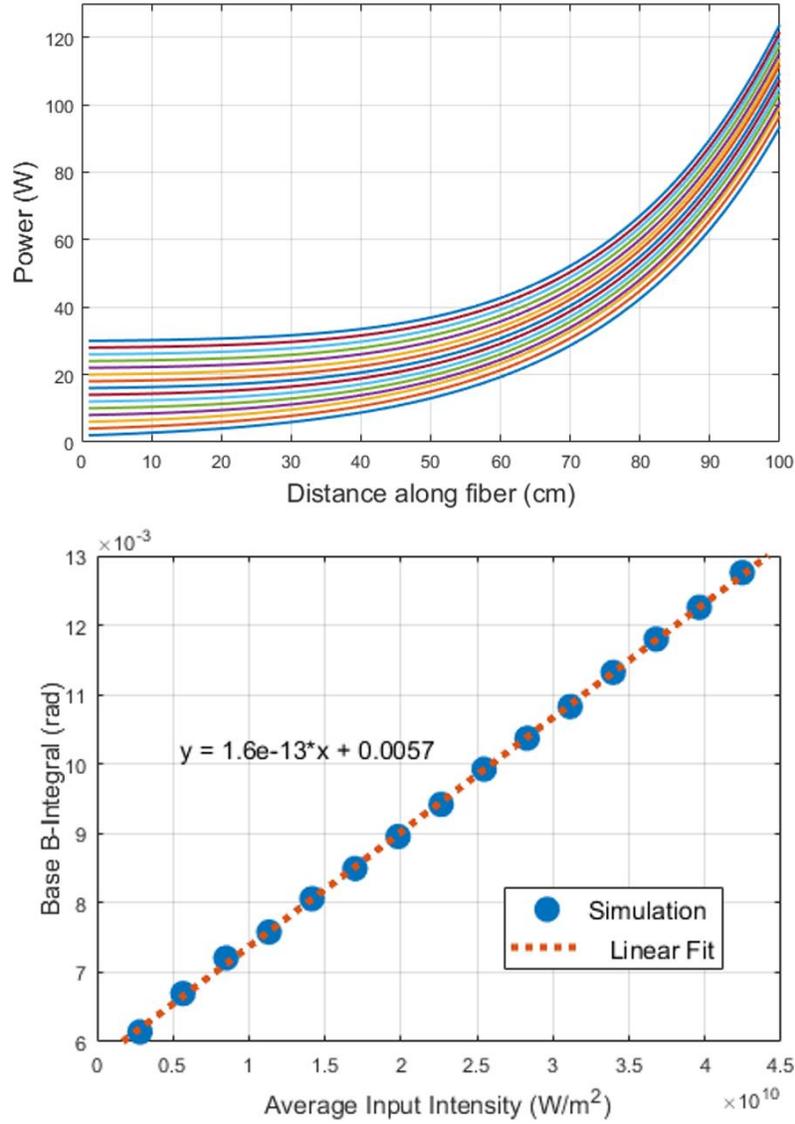

Fig. 7: (Top) Simulated power evolution in an Yb-doped fiber amplifier seeded with average powers of 2 to 30 Watts. (Bottom) B-integral variation with respect to average seed intensity in the simulated fiber amplifier.

With Eq. 3, a B-integral variation between cores of an MCF amplifier system seeded with a range of different average power levels can be estimated based on the single-core amplifier model. For example, consider a 10x10 core MCF amplifier with 30-micron diameter cores with seed light distributed by two 3-Coating SMS optics (Fig.6b). With a 400 W average power seed source (9.7 kW total amplified power), the highest and lowest average seed powers distributed to the cores will be 7.62 W and 2.15 W, respectively. If a stretched-pulse seed laser with $\tau = 3$ ns and $f_{rep} = 100$ kHz is considered, a maximum B-integral difference of 3.88 rad will exist between these two cores after amplification, as modelled for the single core system in Fig. 7. Prior theoretical analysis predicts that this B-integral difference would result in a combination efficiency between these two beams of only 39% [11]. However, in this case the mean B-integral difference between all amplified beams would be 0.67 rad, and the overall loss due to this difference would be only 2% in combination of all 100 beams. In this specific scenario, the distribution of seed light with a pair of 3-Coating simplified SMS optics may be feasible. However, as mentioned previously, this suitability cannot be universally determined, and system parameters will dramatically affect the expected combining efficiency. For example, reducing the seed pulse duration in this scenario from 3 ns to 1 ns yields an expected overall combination efficiency of only 83.5%. Thus, a system-specific analysis must be made that accounts for parameters such as gain fiber lengths, core sizes, overall gain, seed laser pulse duration/repetition rate, etc. Also note that this simplified analysis does not take into account thermal effects specific to multicore active fibers, which can influence B-integral variation and are the subject of ongoing research [12].

The analysis given above is not intended to provide a comprehensive overview of splitting and combining in an MCF amplifier system, as such an analysis would require significantly more detail and simulation tools which take into account the more complex thermal and laser mechanisms occurring in such systems. It is intended, rather, to demonstrate that, in comparison to the case of coherent combination with the simplified SMS designs presented here, determining the suitability of these designs for seed distribution in amplifier systems is a more complex task. Although in many cases the simplified designs may be suitable, additional schemes for simplification, or alternative beam splitting and homogenization techniques may need to be considered depending on amplifier system parameters. Examples might include polarization-based division schemes or optical density filters, both of which may allow for adjustability and better homogenization of distributed beam powers. These and other systems will be the subject of ongoing investigation.

## 4. Conclusion

A numerical analysis of simplified designs for Segmented Mirror Splitters used in coherent combination and splitting of beam arrays has been presented. Due to the relatively weak effect of amplitude mismatch in the coherent combination process, efficient beam combination can be achieved with a reduced number of optimized coating sections, significantly simplifying the manufacturing process of these optics and potentially extending their operation to much higher beam counts than would be feasible with the standard SMS design. The task of beam-splitting and distribution of seed light in multicore fiber amplifier systems presents a more challenging simplification problem, particularly in the case of ultrafast pulsed MCF amplifiers. However, depending on the specific parameters of an amplifier system, the designs presented here may also be suitable for beam-splitting. The simplified SMS concept is applicable not only to MCF systems but also to coherent combination of linear and 2-dimensional arrays in general. Future related work will focus on continuing to simplify multichannel beam handling and control schemes in order to increase the viability of high beam-count coherently combined systems.

## Funding

This work has been supported by the European Research Council (ERC) under grant no. [670557] "MIMAS" and by the German Research Foundation (DFG) within the International Research Training Group (IRTG) 2101.

## Disclosures

The authors declare no conflicts of interest.